\documentclass[preprints,article,accept,oneauthor]{Definitions/mdpi} 

\firstpage{1} 
\pubvolume{1}
\issuenum{1}
\articlenumber{0}
\pubyear{2026}
\copyrightyear{2026}
\datereceived{ } 
\dateaccepted{ } 
\datepublished{ } 
\pdfoutput=1 

\usepackage[linesnumbered,lined,tworuled,commentsnumbered]{algorithm2e}
\usepackage{siunitx}
\usepackage{listings}

\Title{Relative Jump Distance: a Diagnostic for Nested Sampling}

\Author{Johannes Buchner $^{1}$\orcidA{}}
\AuthorNames{Johannes Buchner}
\corres{Correspondence: \href{mailto:johannes.buchner.acad@gmx.com}{johannes.buchner.acad@gmx.com}}
\address{$^{1}$ \quad Max Planck Institute for Extraterrestrial Physics, Giessenbachstrasse, 85748 Garching, Germany}

\abstract{
   Nested sampling is widely used in astrophysics and cosmology to reliably infer model parameters and compare models within a Bayesian framework. To address models with many parameters, Markov Chain Monte Carlo (MCMC) random walks are used within nested sampling to advance a live point population. Diagnostics for nested sampling inference are essential to verify that trustworthy astrophysical conclusions can be drawn.
   We develop a diagnostic to identify poorly mixed random walks that do not meet the requirements of nested sampling.
   The distance from the start to the end of the random walk, the jump distance, is compared to the typical neighbour distance between live points, to obtain a relative jump distance (RJD).
   A robust estimate of a typical neighbour Mahalanobis distance is obtained with the MLFriends algorithm.
   In mock and real-world inference applications such as inferring the distance to GW170817, we observe the relative jump distance. 
   We propose the geometric mean <RJD> and fraction of RJD>1 as new summary diagnostics. 
   To evaluate the sensitivity of the new diagnostics, problematic runs are identified by comparison with reruns using much longer MCMC chains.
   Trustworthy inference runs are consistently associated with <RJD> above 1.1, while problematic runs can fall below these thresholds.
   The RJD is more sensitive than previous tests based on the live point insertion order.
   The RJD diagnostic is proposed as a widely applicable diagnostic to verify inference with nested sampling. It is implemented in the UltraNest package in version 4.1.
}

\keyword{Nested Sampling; Markov Chain Monte Carlo; Bayesian inference; gravitational waves; parameter estimation}

\begin{document}
\SetKwFor{Loop}{Loop:}{}{}

\definecolor{codegreen}{rgb}{0.5,0.5,0.5}
\definecolor{codegray}{rgb}{0.5,0.5,0.5}
\definecolor{codepurple}{rgb}{0.29,0,0.41}
\definecolor{backcolour}{rgb}{0.99,0.99,0.98}
\lstdefinestyle{mystyle}{
    backgroundcolor=\color{backcolour},   
    commentstyle=\color{codegreen},
    keywordstyle=\color{magenta},
    numberstyle=\tiny\color{codegray},
    stringstyle=\color{codepurple},
    basicstyle=\ttfamily\footnotesize,
    breakatwhitespace=false,         
    breaklines=true,                 
    captionpos=b,                    
    keepspaces=true,                 
    numbers=left,                    
    numbersep=5pt,                  
    showspaces=false,                
    showstringspaces=false,
    showtabs=false,                  
    tabsize=2
}
\lstset{style=mystyle}
\section{Introduction}

Nested sampling \citep{Skilling2004,Ashton2022} is a widely used algorithm in astrophysics and cosmology to infer model parameters and compare models within a Bayesian framework. 
To illustrate, the popular packages \citep{multinest,POLYCHORD,DYNESTY,ultranest} have over 6000 citations\footnote{
\href{https://scixplorer.org/search?p=1\&q=citations\%28identifier\%3A\%222008MNRAS.384..449F\%22+OR+identifier\%3A\%222014A\%26A...564A.125B\%22+OR+identifier\%3A\%222015MNRAS.450L..61H\%22+OR+identifier\%3A\%222015MNRAS.453.4384H\%22+OR+identifier\%3A\%222021JOSS....6.3001B\%22+OR+identifier\%3A\%222008MNRAS.384..449F\%22+OR+identifier\%3A\%222020MNRAS.493.3132S\%22+OR+identifier\%3A\%222009MNRAS.398.1601F\%22\%29+property\%3Arefereed}{SciX query}
 retrieved Oct 15, 2026
}, supporting a wide range of astrophysical and cosmological analyses. One appeal of nested sampling is its robustness to complex parameter degeneracies, frequently encountered with sparse or limited observational data common in astrophysics.

Nested sampling achieves this by sampling a population of $K$ live points from the prior probability distribution. Then, at each iteration, a live point is discarded and replaced with a new prior-sample, under the constraint that its likelihood must exceed the likelihood of the recently discarded point. This induces the behaviour of the live point population being distributed throughout the parameter space, but progressing towards the best fit. The key insight of \cite{Skilling2004} is that each iteration of nested sampling discards a constant fraction of the prior probability mass at each iteration, $\delta\approx1/K$.
The posterior distribution can be approximated by the discarded points, weighed by $w_i=L_i\times V_i$, where $V_i=\delta\times(1-\delta)^i$ is the approximated prior probability discarded at iteration $i$, and $L_i$ is the likelihood of the discarded point. Then, the marginal likelihood is $Z\approx\sum w_i$. 
For a more detailed introduction to nested sampling, see \cite{Ashton2022}.
The classic error estimate considers the $V_i$ error, while \cite{Higson2019} proposed bootstrapping of the live points to sample also the variance on $L_i$. This Z error estimate is adopted here.

Several approaches have been proposed for prior-sampling under the likelihood constraint. The original proposal by \cite{Skilling2004} is to start a random walk from a live point for a number of steps $M$, and adopt the final point as a new live point. Points are accepted if they exceed the current likelihood threshold.
Such step sampling nested sampling is the most efficient approach for models with many parameters (say, 20 or more). Slice sampling \cite{neal2003} was first adopted for nested sampling by \cite{jasa2005using} and popularized by \textit{PolyChord} \citep{POLYCHORD}. Various slice sampling approaches were compared by \cite{Buchner2022}.

It is still unclear however how to choose the number of steps $M$. On the one hand, higher $M$ induce a higher computational cost, which may be prohibitive. 
On the other hand, $M$ must be chosen large enough so that the final point is sufficiently independent from the starting point \cite[see also][]{Salomone2018}. 
If $M$ is too small, the inference result and thus scientific conclusions may be wrong. It is thus essential to recognize such problematic nested sampling runs with diagnostics.

Diagnostics for nested sampling can be divided into two categories: visualisations, such as trace plots \citep{Higson2019}, and tests, which this paper focuses on. If the true answer is known, such as in analytic test distributions, the prior mass shrinkage can be tested directly \citep{Buchner2014stats}, as well as the final evidence and posterior samples.
Typically however the true answer is not known.
Here, we can distinguish tests that are generic or specific to the likelihood-restricted prior sampling technique.
For generic tests, consistency of $Z$ across reruns \cite[e.g.][]{Feroz2013,Higson2019} is a basic check. However, consistency does not ensure a good run.
\cite{Fowlie2020} recognized that a newly sampled live point should be inserted into to likelihood-ordered list of live points in a random place. 
The distribution of insertion orders should be uniform, which is checked with a K-S test. \cite{Buchner2021c} generalised this to the integer-based U-test, also applicable when the number of live points is changing during a run \citep[dynamic nested sampling;][]{Higson2017}.
For nested sampling based on rejection sampling, importance nested sampling has been proposed as an alternative estimator \citep{Chopin2010,Feroz2013}. 
Disagreement between the importance nested sampling estimator and the classic nested sampling estimator is a useful indicator with the sampling (see e.g., \citet{Feroz2013,Nelson2020}).
For step samplers, 
comparing the $Z$ estimate when doubling $M$ is a powerful diagnostic \citep{Higson2017,Higson2019}. 
A similar effect can be achieved by increasing the number of live points \citep{SnowballingNS}.
Such reruns with substantially enlarged $M$ or $K$ incur a large computational cost. It would therefore be beneficial to diagnose an inference run by itself, rather than in comparison with other runs. Such a diagnostic is presented in this paper.

\section{Method}

\subsection{Background}
Generalising the idea of \cite{Fowlie2020}, newly sampled points should be unremarkable in all their properties compared to the existing live points. Here, we consider the proximity to the existing live points, and focus in particular on the distance to the live point where the random walk was started. This is motivated by the common failure mode that a walk is not diffusing far away from the starting point.

In nested sampling, the starting point is a randomly chosen live point. Across iterations, therefore, multiple newly sampled points can originate from the same live point. If the walk is not going far, a clump of points near that live point appears. This is then apparent in the distribution posterior sample as graininess of the distribution.
Such graininess has the fundamental property that the distance between properly sampled live points is different than the distance of the random walk. We use this to develop a test.

A performance indicator proposed for Random Walk Metropolis (RWM) is the expected square jump distance (ESJD) \citep[][]{pasarica2010adaptively}. It is defined as the mean square distance between one iteration and the next:
\[ESJD = E[JD^2] = E\left[||\theta_{i+1}-\theta_{i}||^2\right] \]
In particular, \citet{pasarica2010adaptively} consider some univariate variable $\theta$ of interest, such as one model parameter. Then, the square jump distance is simply $(\theta_{i+1}-\theta_{i})^2$.
For a d-dimensional parameter position, the jump distance can be defined as the Euclidean distance. However, a local, decorrelated Mahalanobis distance may be more effective \citep[see e.g. clustering and affine transformations in][]{Higson2017}. In this more general case, with a covariance matrix $\Sigma$, the jump distance (JD) is:
\[JD = \sqrt{\left(\theta_{i+1}-\theta_{i}\right)^T \Sigma^{-1} \left(\theta_{i+1}-\theta_{i}\right)}
\]

To illustrate the behaviour of the ESJD, consider a RWM with a Gaussian proposal kernel.
For a Gaussian proposal with a very small size, the ESJD is low. If the proposal kernel is too large, the Metropolis rule rejects most proposals and the chain remains stuck, again making the ESJD low. Based on this, \cite{pasarica2010adaptively} optimize the kernel scale to maximize the ESJD.
The ESJD is inversely related to the first-order auto-correlation of the random walk \citep{pasarica2010adaptively}, which should be minimized for effective diffusion far away from the starting point. This inverse relationship means that maximising the jump distance is equivalent to minimising the autocorrelation time of the Markov chain, which is the standard criterion for efficient mixing. In the context of nested sampling, where the chain must produce a new live point that is effectively independent of the starting live point, good mixing (large jump distance relative to the live point distribution) is therefore a necessary condition for the validity of the step-sampling approach, even though it is not on its own sufficient to guarantee a correct result.

In the context of nested sampling, \cite{Salomone2018} suggested 
optimizing the expected square jump distance (ESJD) with trial walks. Trials would need to be conducted at each nested sampling iteration, because the distribution is changing with each nested sampling iteration.
Specifically, the prior volume becomes exponentially smaller.

\subsection{Relative Jump Distance}

Here, we develop a new diagnostic by relating the jump distance to the typical distance between live points. 
This normalisation is essential for nested sampling, because the sampled geometry, constrained by the evolving likelihood constraint, is changing in each iteration: early in the run the live points span nearly the full prior volume, while late in the run they may be confined to a tiny region of parameter space. Without normalisation, a raw jump distance that appears large in absolute terms may in fact be small relative to the current scale of the live point distribution, and vice versa. By dividing the jump distance by a robust estimate of the typical inter-point separation, the relative jump distance provides a scale-invariant measure of chain mixing that is directly comparable across all stages of a nested sampling run and across problems of different dimensionality.

We define a reference distance to put the JD in context. The reference distance should be a typical distance between live points. Following other works, for the numerical experiments in this paper, 
we work in a coordinate system defined in natural prior probability units (unit hyper-cube), 
and apply clustering and an affine whitening transformation that makes the live point sample covariance the unit diagonal matrix \citep[see e.g.,][]{Handley2015,Buchner2021UltraNest,Buchner2022}.
This already provides a first normalisation of the jump distances. This is however insufficient, because the sample covariance is not equal to the covariance of the likelihood-restricted prior, and the likelihood-restricted prior may be more complex than an ellipsoidal distribution.
To address the structure of the likelihood-restricted prior contour and its dimensionality-dependent edge behaviour, one can consider several choices, such as the typical nearest neighbor distance of the live points.
However, complex likelihood-restricted prior geometries can make the nearest neighbor distance distribution multi-modal and therefore the mean and median nearest neighbour distance are unstable.

For a robust computation, we adopt as the reference distance the radius computed by the MLFriends algorithm \citep{Buchner2014,Buchner2016}, 
as computed for example by dynesty and UltraNest. 
This is found by splitting the live points into a training sample by bootstrapping (sampling with replacement),
and a complementary test sample of the left-out points, 
and identifying the nearest training point for each test point. 
The MLFriends radius r is then the maximum over all these nearest distances across B bootstrapping rounds. 
Here we adopt B=30. \Cref{sec:ellipsoid-r} presents Python code to compute r. 
A detailed analysis of the MLFriends algorithm is presented in \cite{Buchner2026MLFriends},
but we briefly summarize the behaviour.
The neighbourhood defined by radius r is large enough to include live points that may be absent 
due to the Binomial Point Process sampling of the likelihood-restricted prior, 
and is thus expected to encompass the next live point across B bootstrap rounds. 
Unlike a radius derived purely from the Binomial point process alone, 
which would only account for the average number density of live points, 
r adapts to the local geometry of the prior space, 
making it a meaningful measure of the characteristic distance between live points,
and therefore it is suitable as a reference distance.

For the simplified mono-modal case of ellipsoidally distributed live points, 
\Cref{sec:ellipsoid-r} investigates the dependence of r on dimensionality and the number of live points, and 
finds a Mahalanobis distance of $r\approx 1$ in high dimensions. 
This is likely related to the ``concentration of distances'' effect \citep[see e.g.][for a review]{Francois2007concentration}.

With the reference distance r in hand, we define the relative jump distance as: 
\begin{equation}
RJD=JD/r\label{eq:RJD}
\end{equation}
The threshold RJD=1 has a natural geometric interpretation: 
a jump of distance equal to r means the new point has moved at least as far as the typical nearest-neighbor spacing among live points, 
suggesting it has explored a meaningfully different region of the constrained prior. 
Jumps with RJD<1 indicate the new point lies closer to its starting point than to the nearest other live point, a sign of poor chain mixing.
By definition of the bootstrapped r, multiple live points lie within the radius r. We therefore expect $RJD>1$ for the majority of jumps, if the walk is effective.


\begin{figure}
    \centering
    \includegraphics[width=0.5\textwidth]{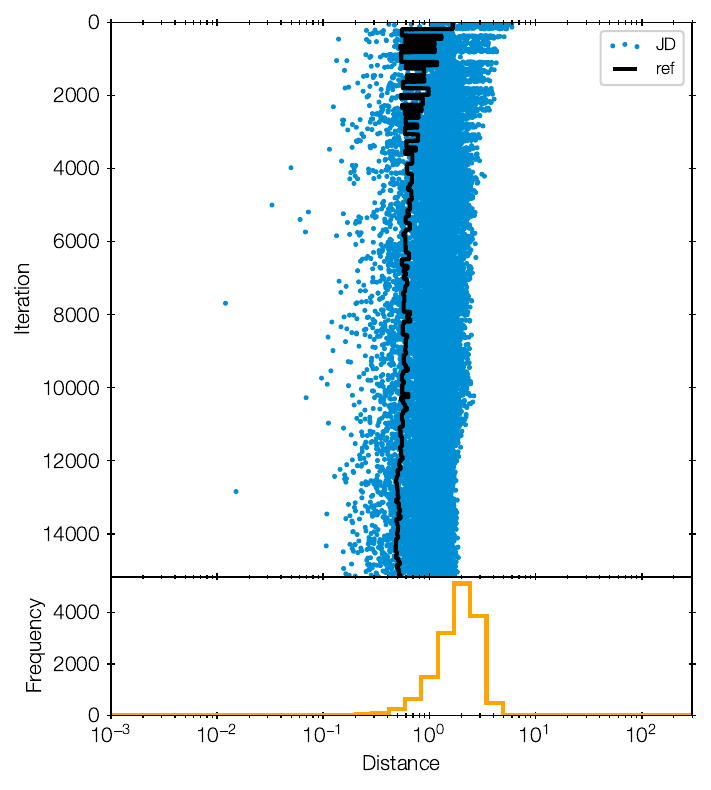}\\
    \caption{Top panel: Jump Distance (blue) and reference Mahalanobis distance (black) at each nested sampling iteration for the run with M=8 steps. Most blue points are to the right of the black curve, i.e., JD>r. Bottom panel: The histogram of the ratio (relative jump distance, JD/r). The majority lies mostly above 1.
    }
    \label{fig:gauss-result-iter}
\end{figure}
\begin{figure}
    \centering
    \includegraphics[width=0.5\textwidth]{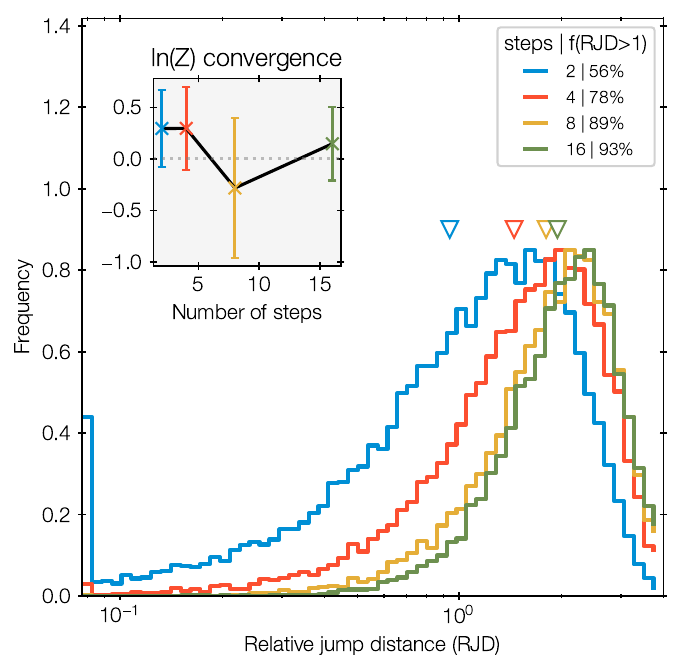}
    \caption{Gaussian test likelihood run diagnostics.
    The inset shows evidence estimates (y-axis) for a sequence of nested sampling runs with different number of steps (x-axis).
    The RJD distribution for each is shown in the main panel.
    The distribution is bell-shaped in logarithmic units, but has a low-end tail, which is more pronounced when the number of steps is smallest (blue). 
    The fraction of RJD>1 is given in the legend with the number of steps.
    The geometric mean of RJDs for each run is shown near the respective histogram with the same color with downward-pointing triangles. 
    Here, the means are all above 1 and the fractions above 50 per cent.
    }
    \label{fig:gauss-result}
\end{figure}

\begin{figure}
    \centering
    \includegraphics[width=0.75\textwidth]{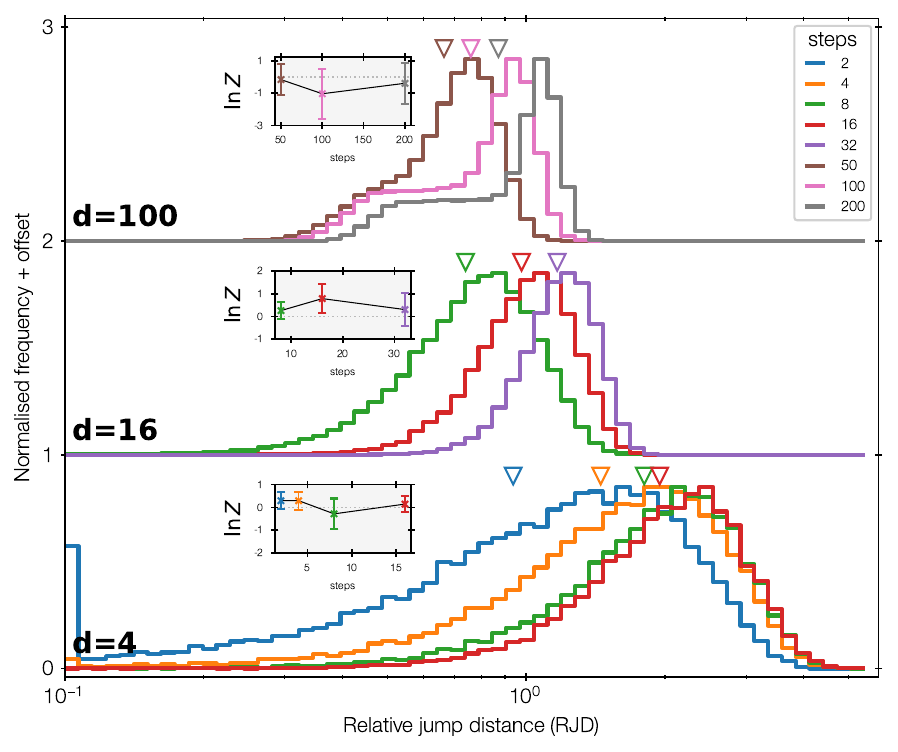}\\
    \caption{The bottom histogram is the same as \cref{fig:gauss-result} for the d=4 Gaussian. This also includes M=2 steps. The offset additional histograms (and corresponding insets) show the d=16 and d=100 case. 
    }
    \label{fig:gauss-result-d}
\end{figure}

\subsection{Summary statistics}

We observe the distribution of RJDs over the nested sampling run. 
To summarize the distribution, we track the fraction of jumps that exceed the reference distance ($f(RJD>1)$) and the geometric mean RJD.
As shown below, the RJD distribution is often log-normal. Problematic runs have a low-RJD tail. 
The geometric mean highlights such a tail better than an arithmetic mean or median, which would be dominated by the unproblematic large RJDs.

The RJD can also be bi-modal. Consider for example a banana geometry in two dimensions. 
If there are enough live points, r describes the typical nearest neighbor distance, insensitive to the length of the banana. 
A good random walk traverses across the banana. A poor walk, with proposals only orthogonal to the banana, is stuck and only traverse approximately the width. 
The fraction of far jumps, $f(RJD>1)$, diagnoses the ratio between good and poor proposals.

\section{Results}

We show the behaviour of our new diagnostic in several applications, which are specified by a $d$-dimensional prior probability distribution and a likelihood function. 
These applications range from standard mathematical test problems to a real gravitational wave parameter estimation problem, demonstrating the relevance of the diagnostic across the spectrum of problems encountered in astrophysics.
For each application, we run nested sampling with 400 live points until the dead points contain more than one per cent of the total estimated posterior weight. 
A sequence of runs is performed with d, $2\times d$, $4\times d$, $8\times d$, etc. number of steps, where d is the number of model parameters. 
For some problems, we start from $d/2$.
For simplicity and to clearly show the behaviour, we use a slice sampler that proposes along one randomly chosen parameter axis.
The seed is fixed across runs.

\subsection{Gaussian}
Integration of a Gaussian likelihood is a standard test problem: 
$L = \prod_{i=1}^{d}\mathrm{Normal}(\mu_{i},\sigma_{i}^{2})$. The true marginal likelihood is $Z\approx1$.
We adopt standard uniform priors with d=4, d=16 and d=100.

The variation adopted here has different standard deviations for each parameter:
$$\sigma_{i} = 10^{\log_{10}(\sigma_\text{min}) + (-1 - \log_{10}(\sigma_\text{min}))\times\frac{i-1}{d-1}}$$
where $\log_{10}(\sigma_\text{min}) = \max\left(-1 - \frac{53}{d}, -7\right)$,
so that $\sigma_{i}$ ranges from $\sigma_\text{min}$ to $10^{-1}$, 
with equal total information $\mathcal{H} = \log(10)\cdot\frac{d}{2}\cdot(-1-\log_{10}(\sigma_\text{min}))$ across dimensionalities where the floor is not active.
Additionally, the means are varied, avoiding a special place (such as the center) of the prior:
$$\mu_{i}=\frac{1}{2}+\frac{1-10\sigma_{i}}{2}\times\sin\frac{i-1}{2}.$$

The jump distances of a nested sampling run with d=4 is shown in 
\Cref{fig:gauss-result-iter}. The JD for each iteration are shown as blue dots and the reference distance r in black, both in whitened coordinates. r is close to 0.5 throughout. The bottom panel shows the RJD distribution. \Cref{fig:gauss-result} shows the RJD distribution for all runs across number of steps. With M=4 steps (blue histogram) the distribution is shifted to the left compared to the histograms with more steps. In all distributions, the majority have RJD>1 (see figure legend), and the geometric mean RJD is also above 1.
The inset in \cref{fig:gauss-result} presents the corresponding ln(Z) estimates from results from runs with different M. These vary around the true value (0). Here both the ln(Z) calibration and the RJD distribution indicate acceptable runs.

\Cref{fig:gauss-result-d} shows a comparison of the d=4, d=16 and d=100 cases.
Here we start from d/2 number of steps. The evidence error bars are 
all overlapping, indicating that the runs do not change with increased number of steps.
Nevertheless, for each case, the histograms change with the number of steps,
with the geometric means moving towards the right in lockstep, and the fraction of RJD>1 increasing. 
The RJD distribution shapes are very different from each other. 
While d=4 and d=16 are bell curves with a strong tail at the low end,
The d=100 case features a plateau. This is likely due to the many initial iterations
where the likelihood-restricted prior still covers the entire unit prior length in
many of the parameters. In these iterations, the geometry is thus similar to a box.

\begin{figure}
    \centering
    \includegraphics[width=0.5\textwidth]{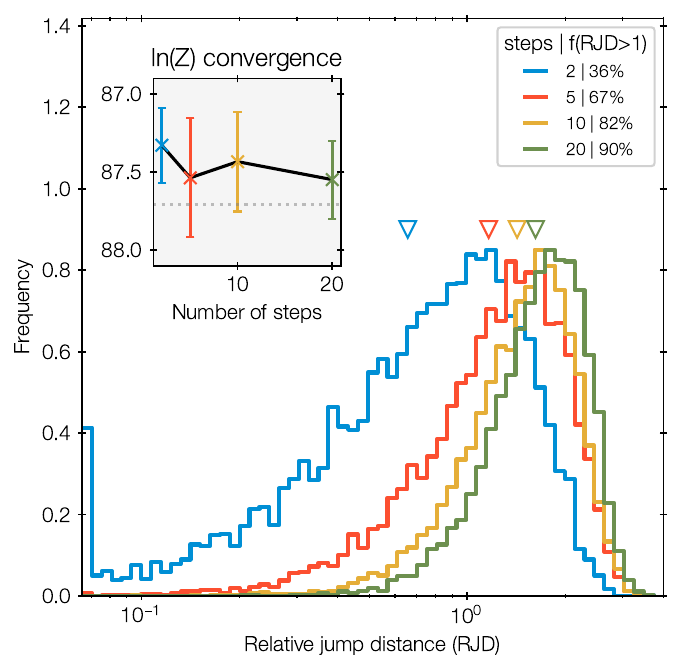}\\
    \caption{Same as \cref{fig:gauss-result}, but for a 5d box.}
    \label{fig:box-result}
\end{figure}

\begin{figure}
    \centering
    \includegraphics[width=0.5\textwidth]{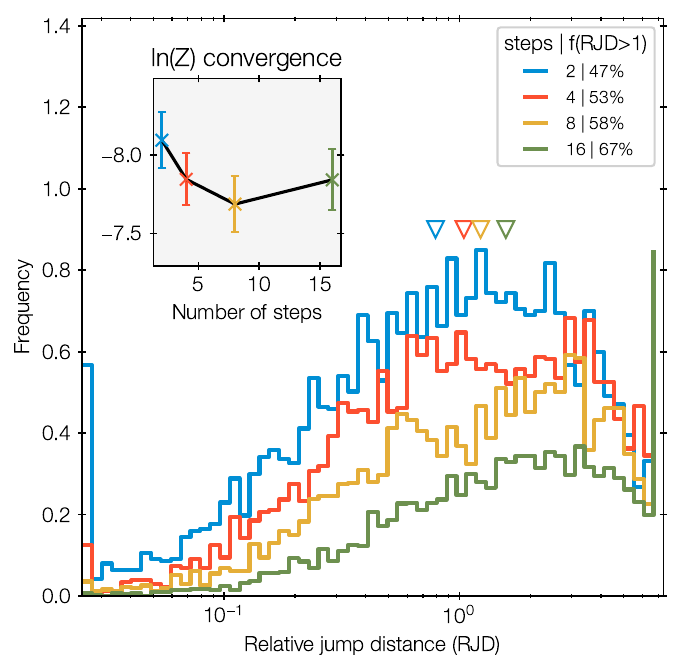}%
    \caption{Results from the banana-like Rosenbrock likelihood in 2d. Same as \cref{fig:gauss-result}.}
    \label{fig:rosenbrock2-result}
\end{figure}
\begin{figure}
    \centering
    \includegraphics[width=0.5\textwidth]{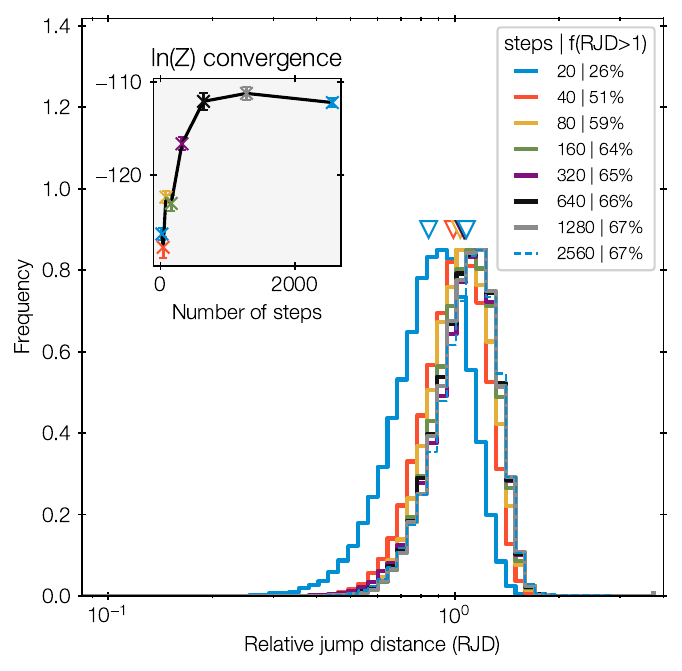}%
    \caption{Same as \cref{fig:rosenbrock2-result}, but for the Rosenbrock likelihood in 20d.
    The RJD distribution (main panel) is bell-shaped in logarithmic units and shifts to the right with increased number of steps.
    The evidence estimate (inset) increases with the number steps until M=640.
    }
    \label{fig:rosenbrock20-result}
\end{figure}

\begin{figure}
    \centering
    \includegraphics[width=\textwidth]{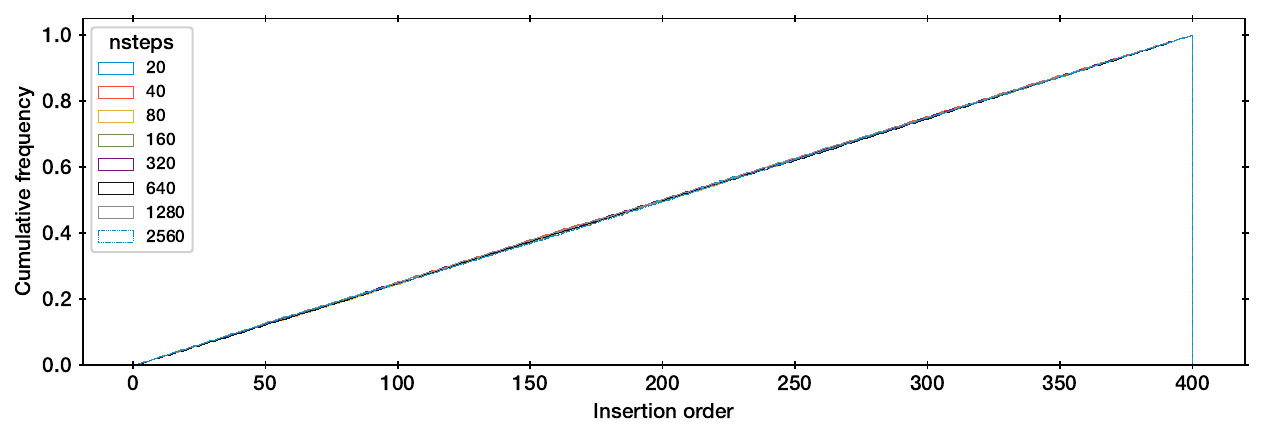}
    \caption{Insertion order distribution for Rosenbrock runs in \cref{fig:rosenbrock20-result}.
    In all cases, the deviation from a uniform distribution is insignificant (p>0.01).}
    \label{fig:rosenbrock-order-result}
\end{figure}

\begin{figure}
    \centering
    \includegraphics[width=0.5\textwidth]{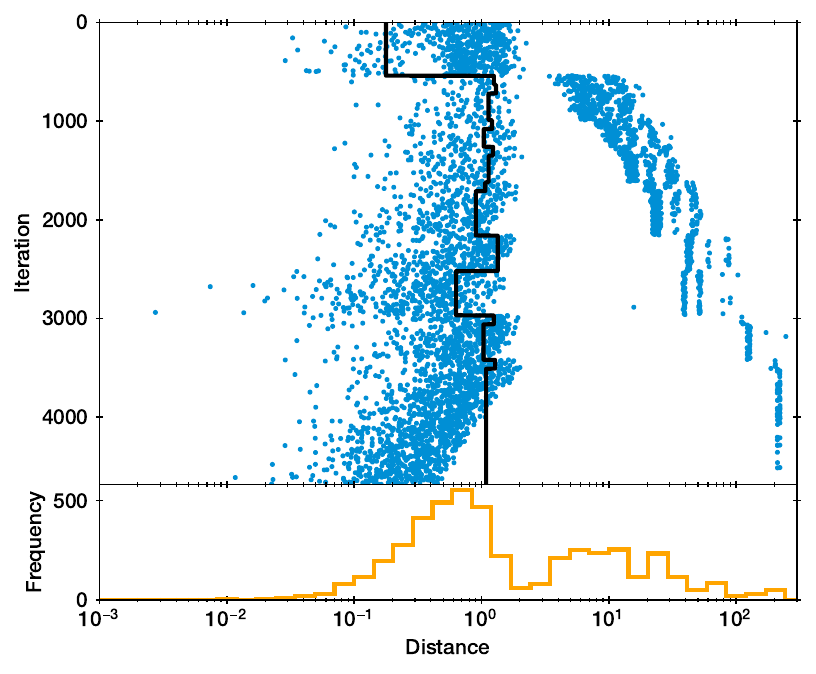}%
    \caption{Same as \cref{fig:gauss-result-iter}, but for the Eggbox likelihood analysed with M=8 steps. 
    The reference distance (black curve) is low in the first 500 iterations, and most jump distances (blue dots) are larger. 
    In the middle, the reference distance is near 1, and the jump distances are either near 1 or above 10. 
    The bottom panel shows the resulting bi-modal RJD distribution.}
    \label{fig:eggbox-result-iter}
\end{figure}
\begin{figure}
    \centering
    \includegraphics[width=0.5\textwidth]{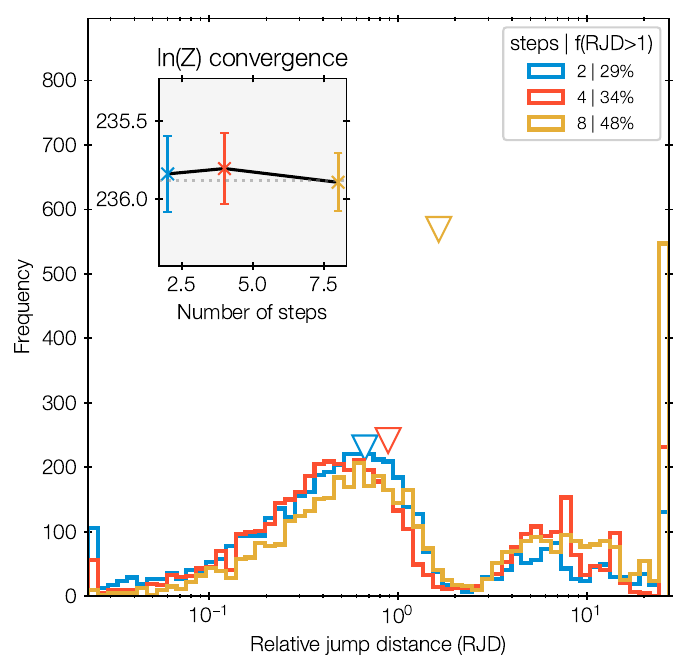}
    \caption{
    The RJD distribution for the eggbox likelihood is bi-modal for all three runs (main panel). The evidence estimates in the inset are consistent within the error bars. }
    \label{fig:eggbox-result}
\end{figure}

\subsection{Box}
We investigate a 5-dimensional non-ellipsoidal likelihood on a standard uniform prior. This is a sharply truncated Gaussian likelihood:
$\ln L= -\frac{1}{2}\times\left(\frac{\theta}{0.1}\right)^{2}+100\times I[\delta<0.1]$, where $\delta=\max_{i}|\theta_{i}|$ is the parameter with the largest deviation from zero. 

The inset of \Cref{fig:box-result} shows little difference in evidence estimates with increasing number of steps. 
The first error bar is however not overlapping with the true value, $\ln Z=87.7$.
However, the histograms of RJD in \Cref{fig:box-result} shift progressively toward higher values the more steps are used. 
With 20 steps, the vast majority of steps are at RJD>1.

\subsection{Rosenbrock function}
We next consider a higher-dimensional, and non-linearly degenerate likelihood.
The Rosenbrock function is a standard test problem in optimization. It exhibits a banana-like degeneracy that can be difficult to navigate. We adopt the formulation
$\log L = -2\times\sum_{i=1}^{d-1}100\times\left(\theta_{i+1}-x_{i}^{2}\right)^{2}+\left(1-\theta_{i}\right)^{2}$
with uniform priors between -10 and +10 for each parameter. 

For d=2, the main panel of \Cref{fig:rosenbrock2-result} shows a mild decrease of evidence estimates with number of steps, and the RJD histograms on the right are comparable. Most RJD are above 1.
The d=20 case is more interesting. The inset of \Cref{fig:rosenbrock20-result} shows a strong change in evidence estimates as the number of steps is increased. Arguably, the result is not converged at fewer than 500 steps.
The RJD histograms on the right also moves to the right with increasing number of steps, however, the effect is subtle from 40-180 steps. The mode of the RJD distribution is just above 1.

To complement these results, we also investigate the effectiveness of the insertion order statistic. 
\Cref{fig:rosenbrock-order-result} presents a cumulative histogram of the insertion orders, which is indistinguishable from a uniform distribution. 
At all number of steps neither the U test nor the KS test raise alarm about the results (p>0.01).

\subsection{Eggbox}
To investigate multi-modality, we adopt the two-dimensional Eggbox likelihood from \cite{Feroz2008}.
It is defined as: 
$\log L	=	\left(2+\cos(5\pi\cdot\theta_{1})\cdot\cos(5\pi\cdot\theta_{2})\right)^{5}$
with uniform priors for each parameter between 0 and $10\pi$.

Here, the observed jump distances have a complex behaviour. \Cref{fig:eggbox-result-iter} shows the jump distances for a nested sampling run. Initially, the reference distance is indicating a typical neighbour distance of the uniformly distributed prior samples. After 500 iterations, the live points concentrate so that clusters can be identified, i.e., r indicates a typical distance within each cluster (r$\sim$1). Some jumps are still possible across clusters (JD>>1), while most remain within (JD$\sim$1). This leads to a bimodal distribution in RJD.
The emerging RJD histograms on the bottom is then bimodal. \Cref{fig:eggbox-result} compares the RJD histograms for runs with different number of steps.
The evidence estimates in the inset of \cref{fig:eggbox-result} show no significant variations. 

\begin{figure}
    \centering
    \includegraphics[width=0.5\textwidth]{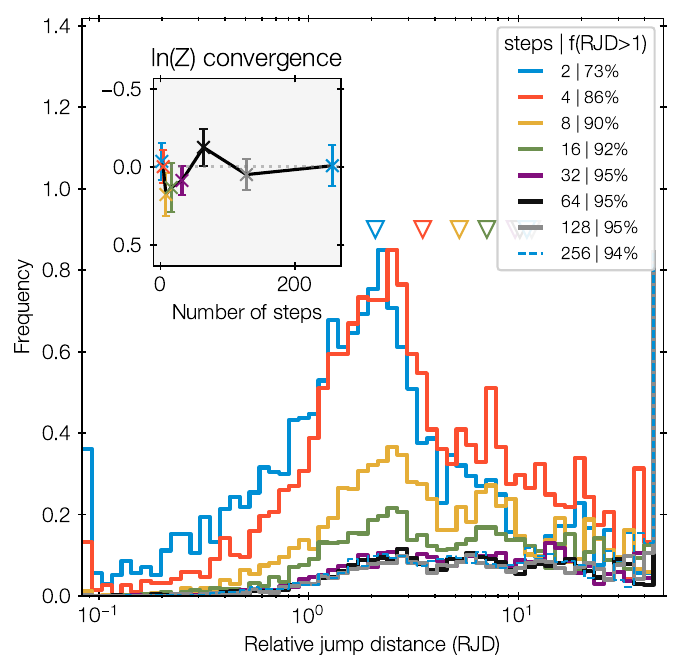}
    \caption{LogGamma likelihood in 2d.
    The RJD distribution (main panel) is bi-modal at high number of steps. The tail at RJD<1 is more pronounced at the lowest number of steps.}
    \label{fig:loggamma2-result}
\end{figure}
\begin{figure}
    \centering
    \includegraphics[width=0.5\textwidth]{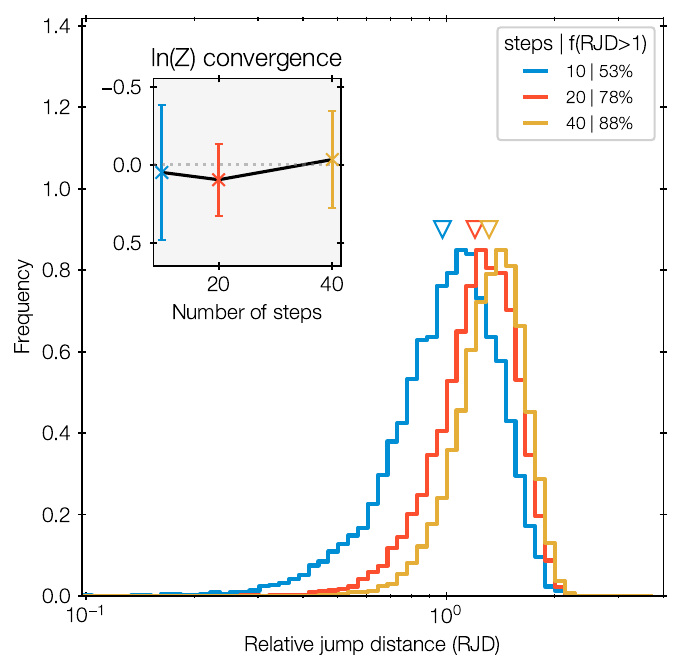}
    \caption{
    The RJD distribution for the LogGamma likelihood in 10d is bell-shaped in logarithmic units and shifts to the right with increased number of steps. 
    The results with 40 and 80 steps are very similar. The evidence estimates in the inset converge to the true value (0) at approximately 40 steps, although all error bars overlap with each other. 
    At 20 steps or higher, the geometric mean indicated by the downward-pointing triangles lies above 1, and the fraction of RJD>1 (listed in the legend) are above 75 per cent.
    }
    \label{fig:loggamma10-result}
\end{figure}

\subsection{LogGamma}
A difficult multi-modal, non-ellipsoidal problem was introduced by \cite{Beaujean2013}, and is defined as:
\begin{align}
g_{a}	\sim	\mathrm{LogGamma}\left(1,\,\frac{1}{3},\,\frac{1}{30}\right);\ & n_{c}	\sim	\mathrm{Normal}\left(\frac{1}{3},\,\frac{1}{30}\right)\\
g_{b}	\sim	\mathrm{LogGamma}\left(1,\,\frac{2}{3},\,\frac{1}{30}\right);\ & n_{d}	\sim	\mathrm{Normal}\left(\frac{2}{3},\,\frac{1}{30}\right)\\
L_{1}	=	\frac{1}{2}\left(g_{a}(x_{1})+g_{b}(x_{1})\right);\ \ \ & L_{2}	=	\frac{1}{2}\left(n_{c}(x_{2})+n_{d}(x_{2})\right)\\
d_{i}	\sim	\mathrm{LogGamma}\left(1,\,\frac{2}{3},\,\frac{1}{30}\right)\ \ \  & \text{if\,\,\,\,\,} 3\leq i\leq\frac{d+2}{2}\\
d_{i}	\sim	\mathrm{Normal}\left(\frac{2}{3},\,\frac{1}{30}\right)\ \ \  & \text{if\,\,\,\,\,} \frac{d+2}{2}<i\\
L	=	L_{1}\times L_{2}\times\prod_{i=3}^{d}d_{i}(x_{i})
\end{align}
We test this problem in 2 and 10 dimensions, with parameters $x_i$ assigned standard uniform priors. The true marginal likelihood is Z=1.

The RJD histograms in \cref{fig:loggamma2-result} is again complex, similar to the Eggbox. Going from the M=2 (blue histogram) and the M=8 case, a tail at low RJD disappears and a hump at $RJD\sim10$ gains prominence. The latter is presumably related to jumps between modes. The evidence estimates are similar (ln(Z)=0.1) and very close to the true value (0). 

At higher dimensions, the evidence estimates are consistent with the true value (0) for runs with 20 steps or more 
(inset of \cref{fig:loggamma10-result}). For 10 steps, the evidence is potentially over-estimated, 
however. Reruns with a different seed gives evidence estimates near zero for all M, indicating that the evidence
trends are not significant.
The main panel of \cref{fig:loggamma2-result} shows that the histogram with 10 steps has the majority of RJD below 1, 
while at higher number of steps, the vast majority are above 1.


\begin{figure}
    \centering
    \includegraphics[width=0.5\textwidth]{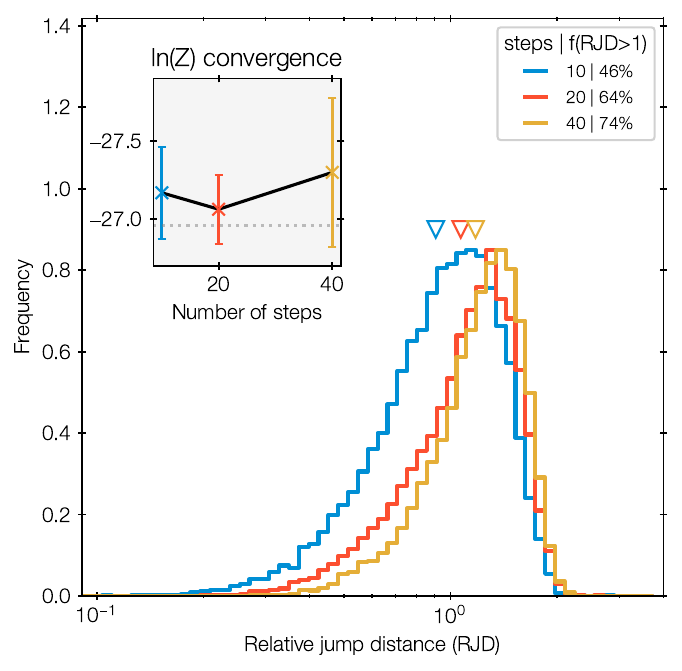}
    \caption{Funnel likelihood in 10d. Same as \cref{fig:gauss-result}.
    }
    \label{fig:funnel-result}
\end{figure}
\begin{figure}
    \centering
    \includegraphics[width=0.75\textwidth]{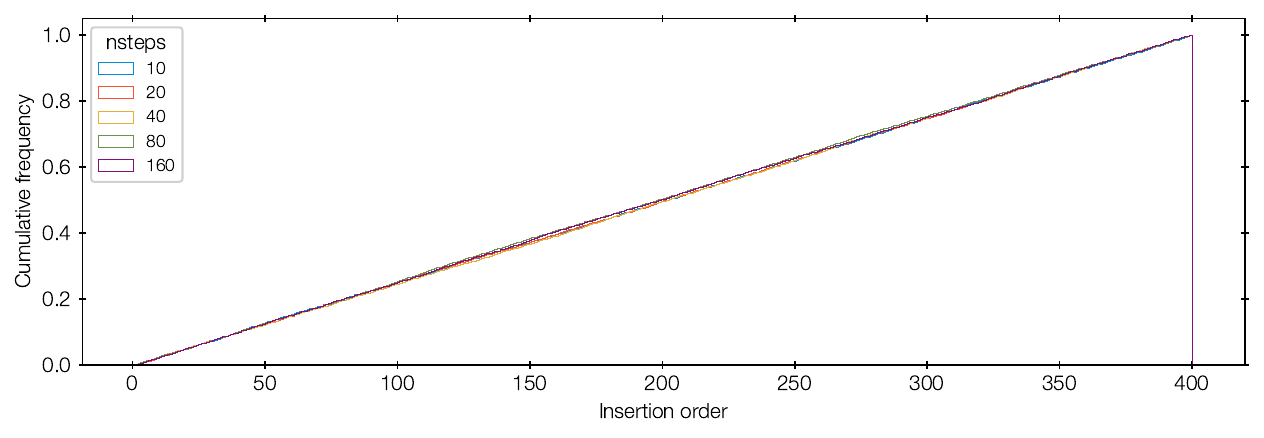}
    \caption{Insertion order distribution for Funnel runs in \cref{fig:funnel-result}.
    In all cases, the deviation from a uniform distribution is insignificant (p>0.01).}
    \label{fig:funnel-order-result}
\end{figure}
\begin{figure}
    \centering
    \includegraphics[width=0.75\textwidth]{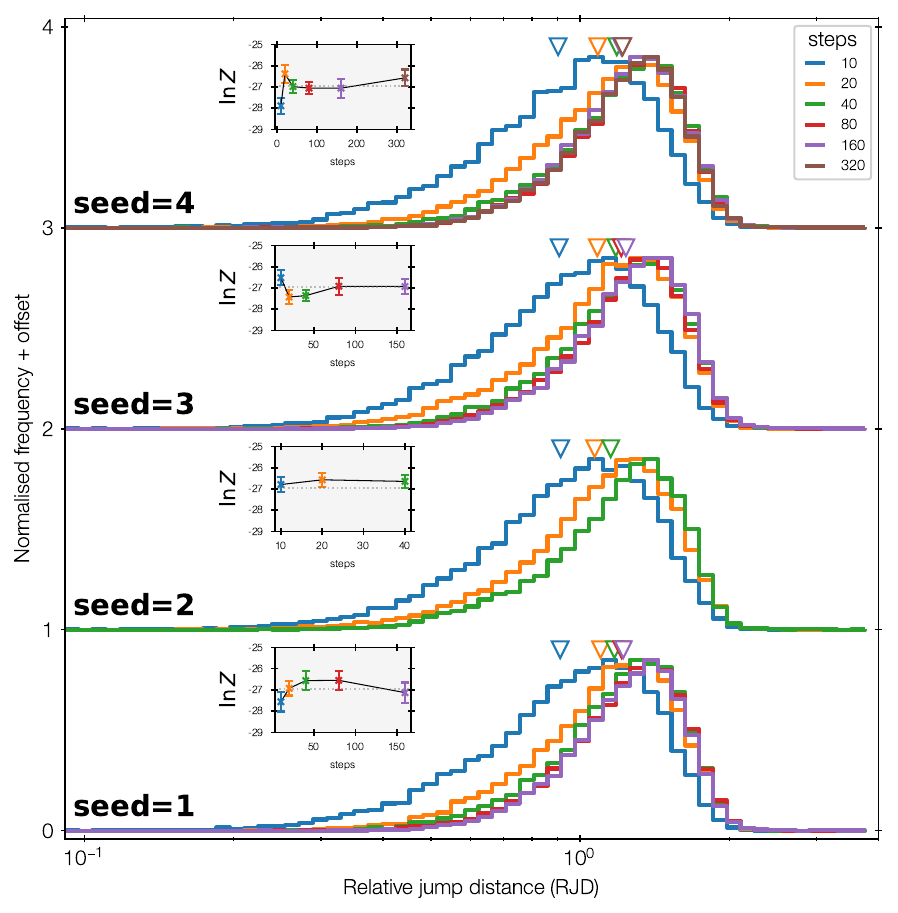}
    \caption{Same as \cref{fig:funnel-result}, but with different random number generator seeds. 
    The number of step doublings until the last three lnZ error bars overlap (see insets) varies. 
    At low number of steps, the evidence estimate scatters within the error bar sizes.
    The RJD distributions of the same color across panels, and 
    geometric means (downward-pointing triangles) are stable.
    }
    \label{fig:funnel-seed-variants}
\end{figure}
\begin{figure}
    \centering
    \includegraphics[width=0.75\textwidth]{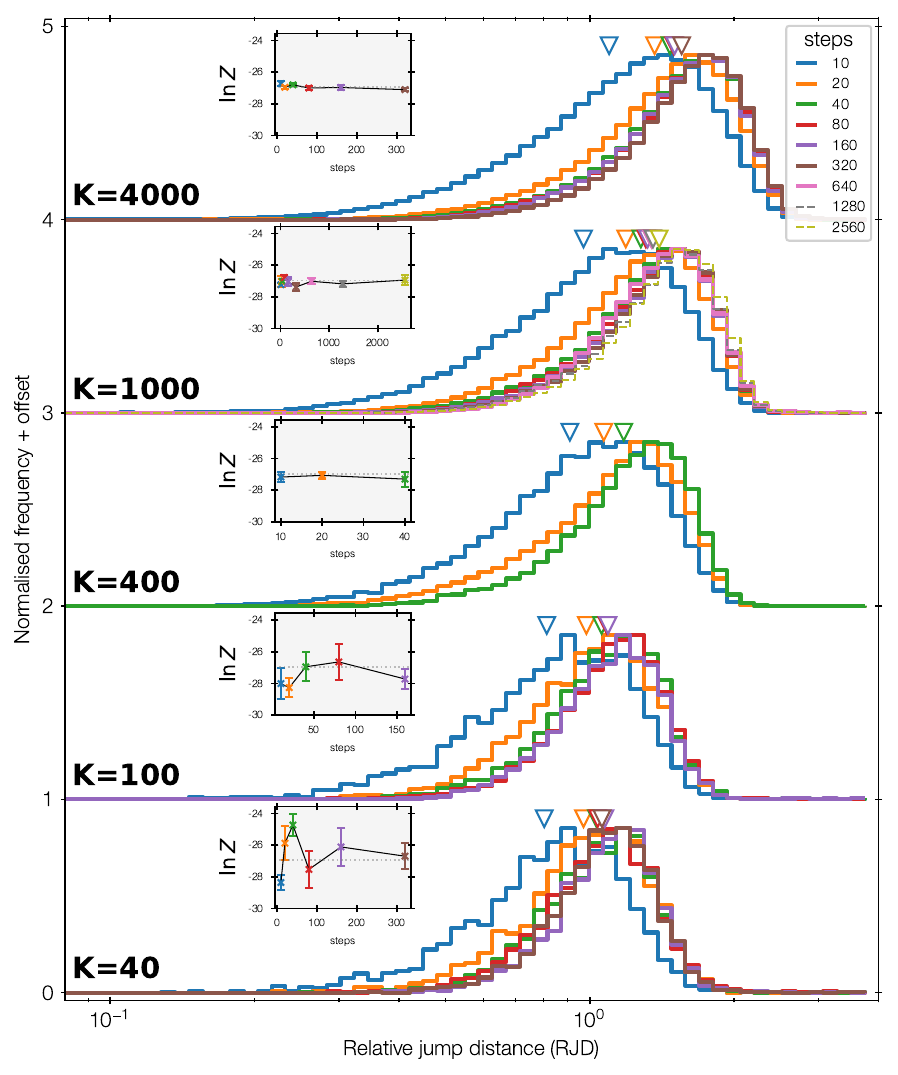}
    \caption{Same as \cref{fig:funnel-result}, but varying the number of live points K.
    Evidence estimates (see insets) scatter more at low K (bottom).
    Comparing the blue histogram from bottom to top rows,
    the RJD distributions at same number of steps (same color) move to the right
    with increasing K.
    }
    \label{fig:funnel-K-variants}
\end{figure}
\begin{figure}
    \centering
    \includegraphics[width=0.75\textwidth]{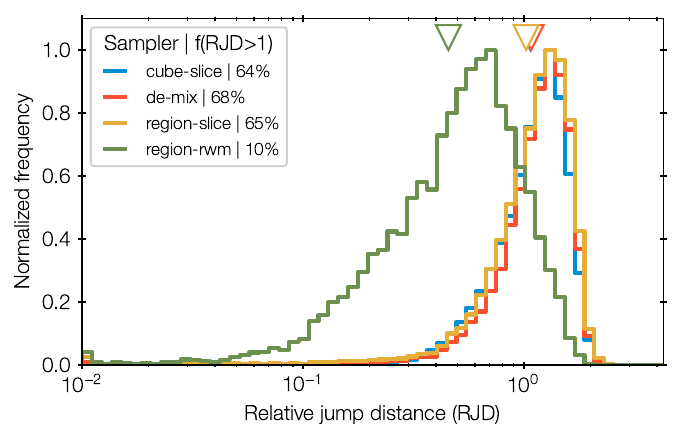}
    \caption{Same as \cref{fig:funnel-result}, but varying the sampling proposal,
    including 
    hypercube axes-aligned slice sampling (cube-slice),
    live point empirical covariance axes-aligned slice sampling (region-harm),
    randomly choosing between region-harm and selecting a direction from a random pair of live points (de-mix),
    Gaussian random walk Metropolis based on empirical live point covariance (region-rwm).
    The RJD distributions of the different slice samplers (blue, red, yellow curves) are very similar,
    while the green curve is at lower typical RJD values.
    }
    \label{fig:funnel-sampler-variants}
\end{figure}

\subsection{Funnel}
\citeauthor{neal2003}'s funnel is a standard test problem that represents features of hierarchical Bayesian models. 
Here we adapt the correlated version of \cite{Karamanis2020} with a correlation matrix $M$ with diagonal elements
$M_{ii}=1$ and off-diagonals set to $M_{ij}=0.95$. 
Then the Gaussian likelihood is 
$$\ln L = -\frac{1}{2} \left[\frac{\mu^{\mathsf T} M^{-1} \mu}{\sigma^2} + d \ln(2\pi\sigma^2) + \ln \det(M) \right]$$
with the standard deviation $\sigma$ and locations $\mu_i$ being model parameters.
The sampled parameters are $\sigma>0$ and scaled deviations $\delta_i=\mu_i/\sigma$, such that $\mu_i=\sigma\delta_i$ 
(noncentered parameterization).
We assign a standard normal prior to $\ln \sigma^2$ and uniform priors between -10 and 10 for each $\mu_i$.
This problem is tested in 10 dimensions (i.e., i from 1 to 9). 
The marginal likelihood, from numerical integration along the $\sigma$ parameter and analytic marginalisation of the other parameters,
is $\ln Z\approx-26.962$.

\Cref{fig:funnel-result} shows that the vast majority of RJD values lie above 1 if more than 10 steps are used.
The evidence estimates in the inset are confined to a narrow range with error bars overlapping with the true value. 
\Cref{fig:funnel-order-result} shows a histogram of the insertion order distribution. 
These are indistinguishable from a uniform distribution for all cases. 
Of these three plots, the RJD distribution appears most sensitive to the number of steps.

We investigate the diagnostic for sensitivity to Monte Carlo random numbers.
\Cref{fig:funnel-seed-variants} shows reruns with different seeds.
Comparing the insets across rows in \cref{fig:funnel-seed-variants} the evidence estimate scatters within the error bar sizes.
Because we double the number of steps until the last three lnZ error bars overlap,
the scatter in the lnZ estimate changes the final number of steps. 
This is reflected in the different number of error bars in the insets
and different number of histograms for each seed row in \cref{fig:funnel-seed-variants}.
Unlike the evidence estimates, the RJD distributions of the same color appears stable across panels.
Consequently, so are the geometric means (downward-pointing triangles).

Next, we investigate the behaviour as th number of live points is changed.
\Cref{fig:funnel-seed-variants} shows five runs with K=40 to 4000 as rows.
As expected, the evidence estimates (see insets) scatter more at low K (bottom),
and overall around the true value (dotted horizontal line).
Comparing the blue histogram in \cref{fig:funnel-seed-variants}  from bottom to top rows,
the RJD distributions at same number of steps (same color) move to the right
with increasing K. While the slice sampler at the same likelihood threshold has
identical behaviour, the spacing of live points, and therefore the reference distance r,
and the distance until a live point is sufficiently decorrelated from the starting point
decreases. This idea is also exploited in snowballing nested sampling \citep{SnowballingNS},
increasing K instead of doubling the number of steps.

Lastly, we test with the same 10d funnel how the diagnostic changes as the proposal is altered.
The results are presented in \cref{fig:funnel-sampler-variants}.
We adopt slice sampling variants previously analysed in \cite{Buchner2022stepsamplers}.
Besides the hypercube axes-aligned slice sampling (cube-slice) used before,
we add the same but proposing along the axis of the empirical covariance of the live points (region-slice).
Next, we add the recommended default in ultranest, which randomly chooses proposals between region-slice 
and selecting a direction from the difference vector of a randomly selected pair of live points (de-mix).
All slice sampling variants have M=20 slice sampling steps with K=400 live points.
The RJD distributions of these different slice samplers (blue, red, yellow curves) are very similar.
These runs cost lie between 1 and 2 million model evaluations.
We also add a RWM, with a multivariate Gaussian proposal distribution based on empirical live point covariance (region-rwm).
The proposed direction sampled is further scaled by a factor, which is, for simplicity, symmetrically multiplied (or divided) on proposal acceptance (rejection) by a small factor $1.1^\frac{1}{M}$.
The meaning of the number of steps M differs here. In slice sampling, M refers to the number of slices proposed and navigated, 
typically requiring O(10) model evaluations. In Random Walk Metropolis, M equals the number of model evaluations.
We chose the region-rwm run with M=160, to obtain a comparable 2 million model evaluations.
The typical RJD values of Random Walk Metropolis, green in \cref{fig:funnel-sampler-variants},
are lower than for slice sampling, indicating insufficient motion of the live points.

\begin{figure}
    \centering
    \includegraphics[width=0.5\textwidth]{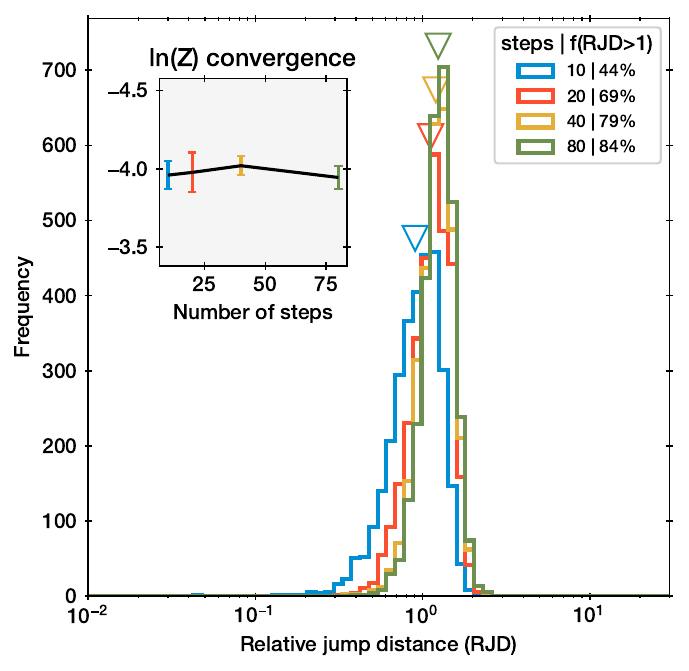}\\
    \caption{The RJD distributions of the eight schools inference runs. These shift to the right with increasing number of steps.
    The evidence estimates shown in the inset overlap within the error bars.
    }
    \label{fig:eightschools-result}
\end{figure}

\begin{figure}
    \centering
    \includegraphics[width=0.5\textwidth]{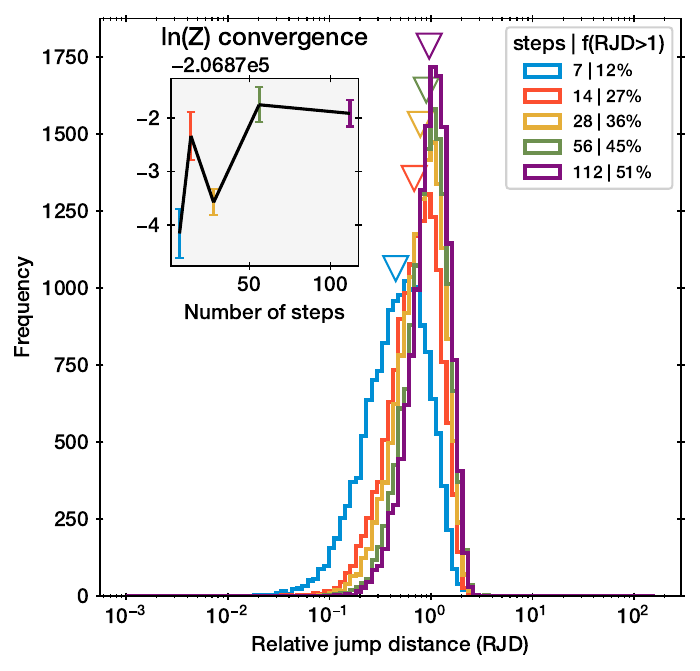}\\
    \caption{The RJD distributions of GW170817 inference runs. These shift to the right with increasing number of steps.
    The evidence estimates are in the inset. They appear not converged at least with fewer than 56 steps.
    The chirp mass mchirp is in solar masses, distance in Mpc, chirp time tc in seconds, while ra and dec are in degrees.
    }
    \label{fig:ligo-result}
\end{figure}
\begin{figure*}
    \centering
    \includegraphics[width=\textwidth]{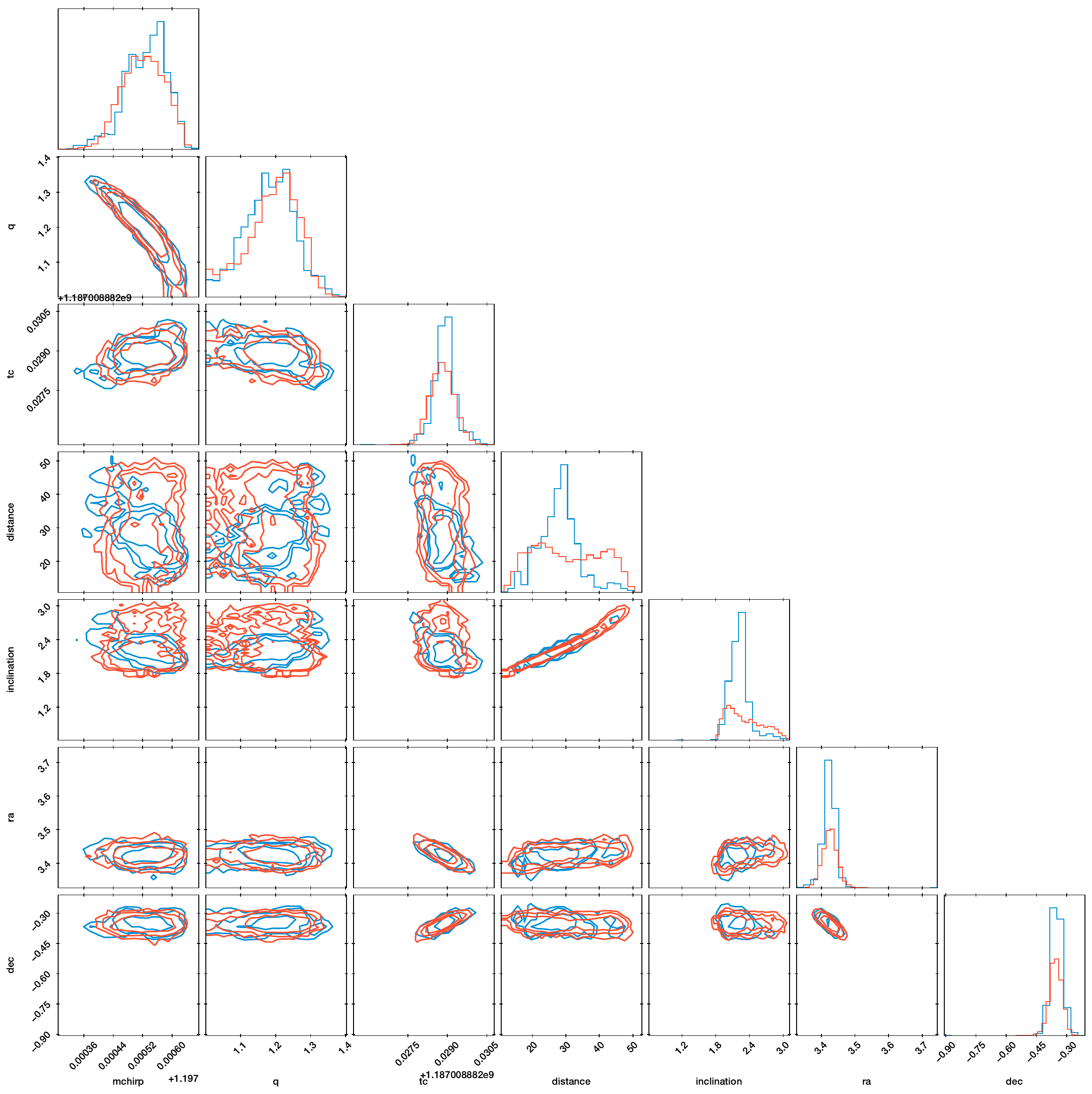}\\
    \caption{Corner plot for GW170817. Red posteriors are from a run with 112 steps, while blue posteriors are from a run with 7 steps.
    The contours include 39.3\%, 86.4\% and 95\% of the posterior probability distribution. Inclination, RA and Dec are given in units of $\pi$.
    }
    \label{fig:ligo-corner}
\end{figure*}

\subsection{Eight schools}
We present a real-world inference problem from \cite{rubin1981estimation}, which obtained Gaussian measurements from eight schools of a treatment effect:
$y\pm\sigma = 28\pm15, 8\pm10, -3\pm16, 7\pm11, -1\pm9, 1\pm11, 18\pm10, 12\pm18$. The mean treatment effect and the variance $\tau^2$ over all schools is of interest.
We therefore define a non-centered parameterization:
$\ln L=\sum_i{\left((x_i \times \tau + \mu) - y_i\right)^2/((2\sigma_i^2)}$
with the x's assigned unit Gaussian priors. The hyper-parameters $\mu$ and $\tau$ are assigned a Gaussian and half-Cauchy distribution, respectively, with mean zero and scale 5.
The results are presented in \cref{fig:eightschools-result}. They are very similar to the Gaussian and Box toy examples. There is no variation in ln(Z), but a noticeable change in RJD. The geometric mean RJD is 0.9 and 1.24 at the lowest and highest number of steps, respectively. The fraction f(JD>r) rises from 44\% to 84\%.

\subsection{Gravitational waves}

The gravitational wave event GW170817 was observed by the LIGO Hanford, Livingston, and Virgo detectors, and inferred to be a binary neutron star inspiral \citep{LIGO2017}.
We follow the PyCBC tutorial\footnote{Tutorial 0 from \href{https://colab.research.google.com/github/gwastro/pycbc-tutorials/blob/master/tutorial/inference\_0\_Overview.ipynb}{https://github.com/gwastro/PyCBC-Tutorials/}} and estimate the masses, inclination, distance, chirp time and position on the sky simultaneously with the marginalized phase Gaussian Noise likelihood \citep[see][]{Biwer2019PyCBCinference} implemented in \cite{pycbcv233}.
For the sine of the inclination angle, a uniform prior is assumed. The prior for the position is uniform on the sphere. The prior on the distance is assumed to be uniform between 10 and 100 Mpc. 
The prior on the chirp mass (in solar masses) and mass ratio q is assumed uniform between 1 and 2. 
The prior on the time of coalescence is assumed uniform between 0.02 and 0.05s of the merger data set time stamp.

\Cref{fig:ligo-result} presents the RJD distribution of a sequence of runs. In the run with few steps (7; blue) the geometric mean lies well below 1 and the f(RJD>1) is very low at 12\%. Subsequent doubling of the number of steps increases the fraction up to 50\%, with the geometric mean of RJD nearing 1. Runs with even more steps were not computed due to the computational cost.

The posterior distribution is presented in \cref{fig:ligo-corner}. 
The constraints are comparable to those published in \cite{Finstad2018}, with minor differences likely arising due to the different, broader priors adopted here. With a low number of steps (7; blue), the distance posterior is much narrower than with the largest number of steps (112; red). This may be because the degeneracy between inclination, distance and time of coalescence could not be sufficiently explored. This demonstrates that the calibration of the number of steps can have consequences for the astrophysical interpretation of the event.

\section{Discussion}

We introduce a new, widely applicable diagnostic for nested sampling. 
It is easy to compute for low and high-dimensional inference tasks and 
does not require multiple nested sampling runs. 

\subsection{Sensitivity and reliability}

The diagnostic is more sensitive than previously proposed tests. 
We compared the RJD distribution, the insertion order U-test and ln(Z) convergence on several applications. 
The relative jump diagnostic is much more sensitive than the insertion order U-test. 
This is likely related to using more information, namely d-dimensional spatial coordinates rather than likelihood quantiles.
Insertion indices discard information on the geometry the likelihood-restricted prior,
so a replacement live point can have an appropriate index while remaining very close to the starting point. 
For step-based sampling in particular, a short chain started at a random live point 
can produce a highly correlated replacement, while not leading to the insertion order test failing.
This highlights a limitation of the test in the case of step-based samplers.
The relative jump diagnostic appears to be at least as sensitive as observing ln(Z). 
An important difference is that, rather than requiring a sequence of runs, 
the fraction of RJD>1 provides a sensitive test of a performed or on-going run in isolation.

This sensitivity arises because the RJD diagnostic directly measures the geometric property, 
chain diffusion relative to the constrained prior extent, that is most closely linked to the failure mode of interest: 
a walk that is not sufficiently mixing within the likelihood-restricted prior. 
The insertion order test \citep{Fowlie2020,Buchner2021c}, 
by contrast, is a one-dimensional test based only on likelihood quantiles, discarding spatial information. 
It can fail to detect chains that are stuck but happen to produce new points at correct likelihood quantiles, 
for example, when multiple modes exist at similar likelihoods but the chain cannot traverse between them. 
The ln(Z) comparison across runs with different step counts is a powerful but expensive global test, 
sensitive to accumulated errors in the evidence integral. 
The RJD diagnostic provides a local, iteration-level signal of the same underlying problem, at negligible additional computational cost.

\begin{figure*}
    \centering
    \includegraphics[width=0.49\textwidth]{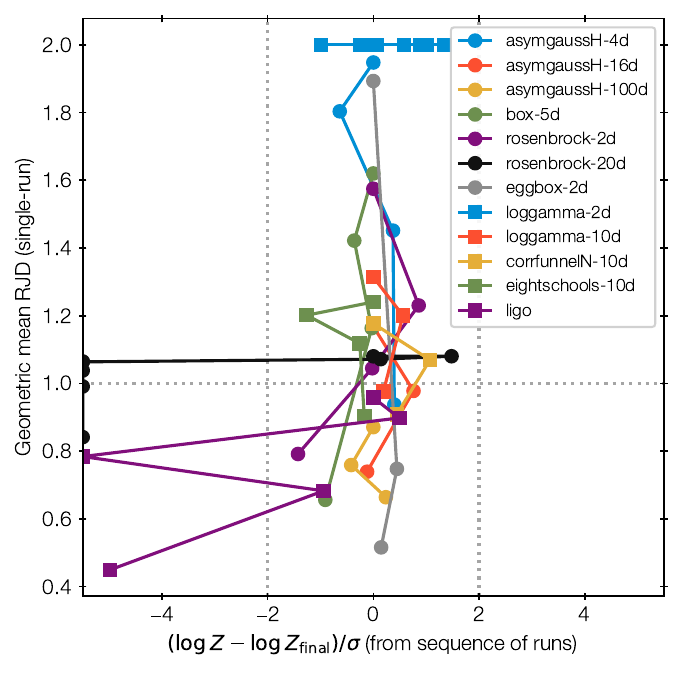}
    \includegraphics[width=0.49\textwidth]{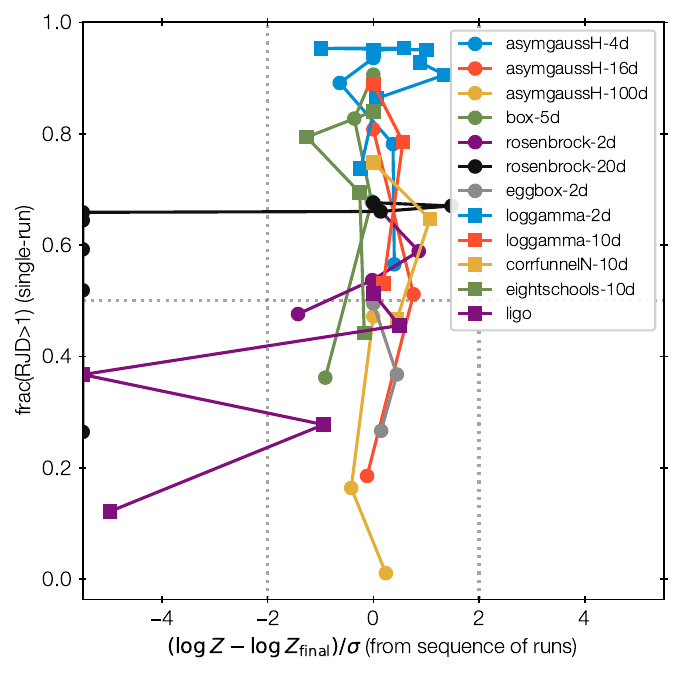}%
    \caption{Collection of all runs, in the relative log-evidence error compared to the 
    geometric mean RJD (left) and the fraction of RJD>1.
    }
    \label{fig:summary}
\end{figure*}

The derivation and validation with toy problems suggest that in mono-modal posteriors 
the nested sampling run should not be considered trustworthy when f(JD<r) 
is below 50 per cent, and the geometric mean of RJD is below 1. 
Intuitively, the former criterion says that the median jump distance 
is below the typical distance between live points.

Although our sample of twelve inference problems in 2 to 100 dimensions is limited, we can attempt to 
understand the false positive and negative rates of these criteria.
\Cref{fig:summary} presents all runs performed (see \cref{appendix:list} for a table of values). 
The lines connect nested sampling runs on identical problems as the number of steps is doubled,
which we term a ladder here.
Considering each ladder of runs, the closer the
evidence estimate to the reference evidence, the higher the diagnostics.
We observe Spearman $\rho<1$ in 12 of 12 cases with a median of -0.50.
\Cref{fig:summary} plots the relative error in log-evidence is plotted against the two new diagnostic criteria.
In this case, we consider 2 sigma deviations from the true evidence value, or 
if not available, from the evidence value derived in the highest M run, problematic.
The relative error is capped at 5.5.
These limits are indicated as vertical dotted lines.
The horizontal dotted line uses our criteria to identify problematic runs (bottom half)
from acceptable runs (top half).
With these nominal thresholds motivated by theory, the criteria are hardly better than chance 
in their false positive and false negative rates (30-50\%), most notably rosenbrock-20d.
However, with thresholds of geometric mean RJD > 1.1, there are 0 (out of 9) false positives with a false negative rate of 44\% (19/43).
With frac(RJD>1)>70\%, the false positive rate is also 0\% (0/9) with a false negative rate of 56\% (24/43).
Those criteria are therefore primarily useful to vouch for runs, but failing the criterion does not necessarily imply a bad result.
A caveat here is that for the most realistic problems the true evidence value is unknown, 
and proper calibration of thresholds would need a larger sample with a hold-out set.

The expected relative jump distance can be over-estimated. 
If the number of live points is low and/or the dimensionality high, clusters and local structures may be unrecognizable when computing the typical neighbour distance. 
Similarly, if the reference distance is only computed every nth iteration, 
the moment when a likelihood-restricted prior becomes multi-modal and modes can no longer be traversed may be missed. 
At that moment, r may reflect the spacing across the full multi-modal distribution rather than within a single mode, 
causing r to be over-estimated and RJD to appear deceptively low even when the walk is mixing well within each mode.
We observe such behaviour with the multi-modal Eggbox and Loggamma toy problems.

Based on our results, \textbf{we preliminarily recommend the following procedure for practitioners}: 
(1) If the geometric mean RJD is above 1.1, then the nested sampling result might be trustworthy.
(2) Otherwise, rerun with twice as many steps. If either the RJD distribution changed, or the Z estimate changed significantly, go to (1). Otherwise, the result may be acceptable.
The cautious practitioner should double the number of steps once more than seems necessary.
The major computational benefit of this procedure is that any alteration of the Bayesian model 
does not automatically incur a re-calibration of the number of steps.

\subsection{Future work}

In this work, we have primarily focused on the reliability of evidence estimates,
as commonly used for Bayesian model comparison. 
The effect on the posterior distribution may be different, 
as previously discussed for MultiNest reliability by \cite{Dittmann2024}.
Nevertheless, \cref{fig:ligo-corner} shows an example 
where posterior distributions are incorrectly estimated by one run and 
the criteria flags it as unreliable.

The RJD diagnostic compares the new-born live point to the starting point of its random walk. 
This makes it specific to step sampling-based nested samplers. 
A generalization would be to consider the distance of the new-born point to its nearest existing live point, 
rather than the starting point. Such a generalization could be implemented without modification to existing nested samplers. 
Our diagnostic is a stronger form, because it uses the most correlated point. 
Implementation in existing nested sampling packages is easy and demands little additional resources, 
since only the distance between starting and final point needs to be stored. 
The RJD diagnostic is implemented in \textit{UltraNest}\footnote{https://johannesbuchner.github.io/UltraNest/} from version 4.1.

Future work will explore ways to adapt the number of steps during a nested sampling run. 
Preliminary implementations in \textit{UltraNest} are available which increase the number of steps after each iteration where RJD<1, 
and decrease it after each iteration where RJD>1. 
An alternative approach would be to use Snowballing nested sampling \citep{SnowballingNS}, i.e., 
choose a fixed number of steps and bulk up the number of live points in likelihood ranges where f(JD>r) was low. 
As the density of live points increases, the typical neighbour distance r decreases, 
which increases RJD without the need to discard existing chains.

\acknowledgments{
I thank the two anonymous reviewers and the academic editor for comments and suggestions that improved the manuscript.
I thank Will Handley for insightful conversations, which were hosted by the Kavli Institute for Cosmology in Cambridge, UK.
}

\dataavailability{
The code for generating the data is available at
\url{https://github.com/JohannesBuchner/paper-relative-jump-distance}.
}

\useofartificialintelligence{
AI (primarily Anthropic Claude 4.6) was used for improving the grammar and isolated components of the data analysis.
}

\reftitle{References}
\bibliography{ms}

\appendixtitles{no} 
\appendixstart
\appendix
\section[\appendixname~\thesection]{Reference distance for ellipsoids}
\label{sec:ellipsoid-r}
To illustrate the behaviour of the defined reference distance, we can consider ellipsoidal contours. 
If we consider that an ellipsoid volume $V$ that is segmented into K equal sub-volumes with a volume proportional to $r^d$, then $V\propto K\times r^d$. For fixed V, we find the scale between neighbours $r$ scales as $r\propto K^{-\frac{1}{d}}$.


\begin{figure}
    \centering
    \includegraphics[width=0.8\textwidth]{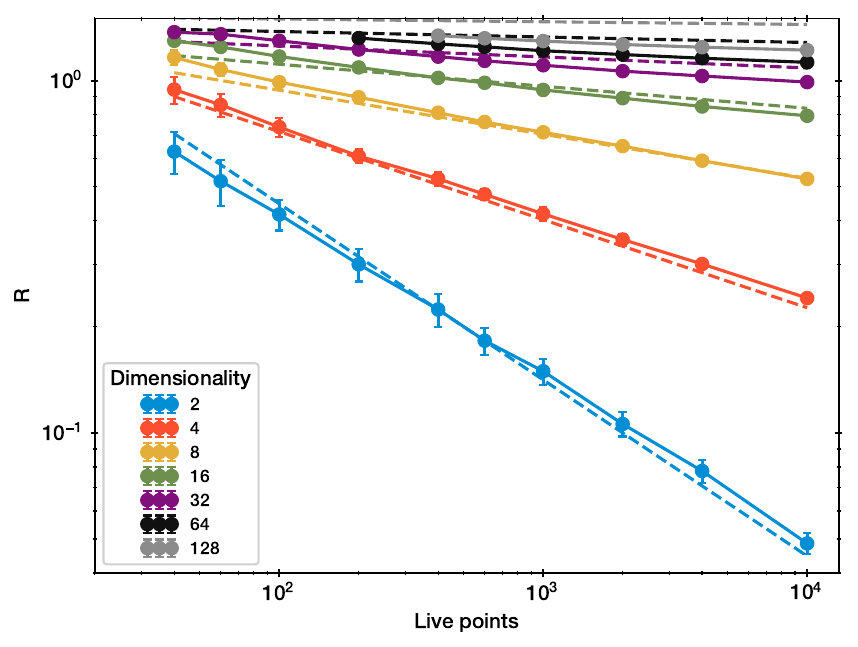}
    \caption{Scaling of reference radius with number of live points and dimensionality. Each error bar summarizes the mean and standard deviation from 40 simulations. The dashed curves is \cref{eq:approx-scaling-ellipsoids}. This approximates most data points, but there are deviations in high dimensions and low number of live points.}
    \label{fig:scaling-ellipsoids}
\end{figure}

We perform numerical experimentation sampling K points from an ellipsoid in up to 128 dimensions. 
We compute r with the code listed in \cref{fig:computeradius}.
The found r values are presented in \cref{fig:scaling-ellipsoids}. 
We also plot the relation: 
\begin{equation}
r = (20/K)^\frac{1}{d} \times (d/2)^{0.1}
\label{eq:approx-scaling-ellipsoids}
\end{equation}
As discussed above, the first term is the volumetric distribution by the K live point in a d-dimensional hyper-sphere. The weak second term is likely related to the empirical whitening transformation failing to accurately identify the true whitening transform in high dimensions. 

\begin{figure}
\caption{Python code to compute MLFriends radius. Function \texttt{compute\_radius} is called with an array of live points.}
\begin{lstlisting}[language=Python,numbers=none]
from ultranest.mlfriends import MLFriends, AffineLayer
def compute_radius(live_points):
    # start with a Euclidean distance
    layer = AffineLayer()
    # run MLFriends, find radius
    region = MLFriends(live_points, layer)
    r, _ = region.compute_enlargement()  
    # Create Mahalanobis distance, considering clusters within distance r 
    layer2 = layer.create_new(live_points, r)
    # run MLFriends and return radius
    region2 = MLFriends(live_points, layer2)
    r2, _ = region2.compute_enlargement()
    return r2**0.5
\end{lstlisting}
    \label{fig:computeradius}
\end{figure}

In high dimensions we find that $r$ lies between  1.2 and 1.3, independent of the number of live points. Likely this is related to the number of bootstrapping rounds. $r=1$ intuitively corresponds to the Mahalanobis distance of any ellipsoid axis. If we were to adopt $r=1$, the RJD criterion would become more optimistic about the achieved jump length.  
In more realistic settings than the ellipsoidal contours considered here, $r$ may be substantially different.

\section[\appendixname~\thesection]{Listing of inference runs}
\label{appendix:list}
A list of the runs presented in this work is given in \Cref{tab:summary,tab:summaryfunnel,tab:summaryfunnelseed}.

\begin{table}
    \centering
    \caption{Summary of runs: name, dimensionality ($d$), number of live points ($N_\mathrm{live}$), number of steps, geometric mean RJD, fraction with RJD$>1$, evidence estimate $\ln Z \pm \sigma$, and true value (if available). 
    Bold geometric mean RJD indicates $<1$; bold $f(\mathrm{RJD}>1)$ indicates $<0.5$. All listed here are run with K=400 and the unit-slice proposal.}
    \label{tab:summary}
\begin{tabular}{lrrrrrrrl}
\hline
Name & $d$ & Steps & <RJD> & $f(\mathrm{RJD}>1)$ & $\ln Z$ & $\sigma_{\ln Z}$ & $\ln Z_\mathrm{true}$ \\
\hline
asymgaussH-4d & 4 & 2 & \textbf{0.938} & 0.565 & 0.30 & 0.37 & 0.00 \\
 & 4 & 4 & 1.451 & 0.782 & 0.30 & 0.40 & 0.00 \\
 & 4 & 8 & 1.804 & 0.891 & -0.28 & 0.68 & 0.00 \\
 & 4 & 16 & 1.948 & 0.936 & 0.15 & 0.36 & 0.00 \\
asymgaussH-16d & 16 & 8 & \textbf{0.739} & \textbf{0.185} & 0.26 & 0.39 & 0.00 \\
 & 16 & 16 & \textbf{0.978} & 0.512 & 0.78 & 0.63 & 0.00 \\
 & 16 & 32 & 1.168 & 0.808 & 0.30 & 0.72 & 0.00 \\
asymgaussH-100d & 100 & 50 & \textbf{0.664} & \textbf{0.011} & -0.15 & 0.96 & 0.00 \\
 & 100 & 100 & \textbf{0.759} & \textbf{0.164} & -1.03 & 1.55 & 0.00 \\
 & 100 & 200 & \textbf{0.871} & \textbf{0.472} & -0.38 & 1.28 & 0.00 \\
\hline
box & 5 & 2 & \textbf{0.656} & \textbf{0.362} & 87.33 & 0.24 & 87.71 \\
 & 5 & 5 & 1.163 & 0.678 & 87.54 & 0.38 & 87.71 \\
 & 5 & 10 & 1.422 & 0.827 & 87.43 & 0.32 & 87.71 \\
 & 5 & 20 & 1.620 & 0.906 & 87.55 & 0.25 & 87.71 \\
\hline
rosenbrock-2d & 2 & 2 & \textbf{0.792} & \textbf{0.476} & -8.10 & 0.18 & --- \\
 & 2 & 4 & 1.045 & 0.537 & -7.85 & 0.17 & --- \\
 & 2 & 8 & 1.230 & 0.589 & -7.69 & 0.18 & --- \\
 & 2 & 16 & 1.575 & 0.675 & -7.84 & 0.20 & --- \\
rosenbrock-20d & 20 & 20 & \textbf{0.841} & \textbf{0.264} & -126.35 & 0.65 & --- \\
 & 20 & 40 & \textbf{0.991} & 0.519 & -127.78 & 1.11 & --- \\
 & 20 & 80 & 1.038 & 0.592 & -122.38 & 0.63 & --- \\
 & 20 & 160 & 1.065 & 0.644 & -123.04 & 0.87 & --- \\
 & 20 & 320 & 1.064 & 0.659 & -116.64 & 0.70 & --- \\
 & 20 & 640 & 1.072 & 0.660 & -112.07 & 0.93 & --- \\
 & 20 & 1280 & 1.080 & 0.670 & -111.21 & 0.67 & --- \\
 & 20 & 2560 & 1.080 & 0.676 & -112.20 & 0.49 & --- \\
\hline
eggbox-2d & 2 & 2 & \textbf{0.516} & \textbf{0.266} & 235.79 & 0.24 & 95.08 \\
 & 2 & 4 & \textbf{0.748} & \textbf{0.368} & 235.82 & 0.16 & 95.08 \\
 & 2 & 8 & 1.893 & \textbf{0.495} & 235.75 & 0.18 & 95.08 \\
\hline
loggamma-2d & 2 & 2 & 2.080 & 0.737 & -0.04 & 0.12 & 0.00 \\
 & 2 & 4 & 3.501 & 0.863 & -0.00 & 0.11 & 0.00 \\
 & 2 & 8 & 5.231 & 0.906 & 0.18 & 0.14 & 0.00 \\
 & 2 & 16 & 7.075 & 0.928 & 0.13 & 0.16 & 0.00 \\
 & 2 & 32 & 9.648 & 0.951 & 0.09 & 0.09 & 0.00 \\
 & 2 & 64 & 10.160 & 0.953 & -0.12 & 0.12 & 0.00 \\
 & 2 & 128 & 11.576 & 0.954 & 0.05 & 0.10 & 0.00 \\
 & 2 & 256 & 10.981 & 0.949 & -0.01 & 0.13 & 0.00 \\
loggamma-10d & 10 & 10 & \textbf{0.976} & 0.531 & 0.05 & 0.43 & 0.00 \\
 & 10 & 20 & 1.201 & 0.785 & 0.09 & 0.23 & 0.00 \\
 & 10 & 40 & 1.314 & 0.889 & -0.03 & 0.31 & 0.00 \\
\hline
eightschools-10d & 10 & 10 & \textbf{0.903} & \textbf{0.442} & -3.96 & 0.09 & --- \\
 & 10 & 20 & 1.118 & 0.694 & -3.98 & 0.13 & --- \\
 & 10 & 40 & 1.202 & 0.794 & -4.02 & 0.06 & --- \\
 & 10 & 80 & 1.241 & 0.840 & -3.94 & 0.07 & --- \\
\hline
ligo & 7 & 7 & \textbf{0.449} & \textbf{0.122} & -206874.17 & 0.45 & --- \\
 & 7 & 14 & \textbf{0.683} & \textbf{0.277} & -206872.33 & 0.45 & --- \\
 & 7 & 28 & \textbf{0.784} & \textbf{0.367} & -206873.58 & 0.24 & --- \\
 & 7 & 56 & \textbf{0.898} & \textbf{0.456} & -206871.75 & 0.33 & --- \\
 & 7 & 112 & \textbf{0.958} & 0.514 & -206871.91 & 0.25 & --- \\
\hline
\end{tabular}

\end{table}

\begin{table}
    \centering
    \caption{Summary of runs, continued, for the funnel with d=10 (corrfunnelN-10d).}
    \label{tab:summaryfunnel}
\begin{tabular}{lrrrrrrrl}
\hline
Variant & $K$ & Steps & <RJD> & $f(\mathrm{RJD}>1)$ & $\ln Z$ & $\sigma_{\ln Z}$ & $\ln Z_\mathrm{true}$ \\
\hline
cube-slice-seed42& 40 & 10 & \textbf{0.804} & \textbf{0.307} & -28.36 & 0.49 & -26.96 \\
& 40 & 20 & \textbf{0.970} & 0.506 & -25.88 & 1.06 & -26.96 \\
& 40 & 40 & 1.033 & 0.591 & -24.74 & 0.70 & -26.96 \\
& 40 & 80 & 1.039 & 0.596 & -27.54 & 1.15 & -26.96 \\
& 40 & 160 & 1.075 & 0.656 & -26.12 & 1.20 & -26.96 \\
& 40 & 320 & 1.060 & 0.636 & -26.71 & 0.81 & -26.96 \\
cube-slice-seed42& 100 & 10 & \textbf{0.813} & \textbf{0.341} & -28.01 & 0.96 & -26.96 \\
& 100 & 20 & \textbf{0.984} & 0.528 & -28.26 & 0.62 & -26.96 \\
& 100 & 40 & 1.059 & 0.631 & -26.95 & 0.92 & -26.96 \\
& 100 & 80 & 1.086 & 0.676 & -26.64 & 1.12 & -26.96 \\
& 100 & 160 & 1.091 & 0.667 & -27.72 & 0.64 & -26.96 \\
cube-slice-seed42& 400 & 10 & \textbf{0.909} & \textbf{0.467} & -27.17 & 0.29 & -26.96 \\
& 400 & 20 & 1.070 & 0.647 & -27.06 & 0.22 & -26.96 \\
& 400 & 40 & 1.177 & 0.748 & -27.30 & 0.48 & -26.96 \\
cube-slice-seed42& 1000 & 10 & \textbf{0.971} & 0.536 & -27.22 & 0.13 & -26.96 \\
& 1000 & 20 & 1.189 & 0.731 & -27.02 & 0.33 & -26.96 \\
& 1000 & 40 & 1.278 & 0.802 & -27.01 & 0.16 & -26.96 \\
& 1000 & 80 & 1.309 & 0.822 & -26.78 & 0.15 & -26.96 \\
& 1000 & 160 & 1.321 & 0.833 & -27.04 & 0.27 & -26.96 \\
& 1000 & 320 & 1.348 & 0.852 & -27.37 & 0.27 & -26.96 \\
& 1000 & 640 & 1.345 & 0.854 & -27.01 & 0.19 & -26.96 \\
& 1000 & 1280 & 1.355 & 0.851 & -27.17 & 0.21 & -26.96 \\
& 1000 & 2560 & 1.396 & 0.879 & -26.93 & 0.29 & -26.96 \\
cube-slice-seed42& 4000 & 10 & 1.097 & 0.641 & -26.73 & 0.15 & -26.96 \\
& 4000 & 20 & 1.365 & 0.819 & -26.97 & 0.10 & -26.96 \\
& 4000 & 40 & 1.470 & 0.871 & -26.81 & 0.08 & -26.96 \\
& 4000 & 80 & 1.497 & 0.882 & -27.01 & 0.13 & -26.96 \\
& 4000 & 160 & 1.503 & 0.890 & -26.98 & 0.17 & -26.96 \\
& 4000 & 320 & 1.555 & 0.915 & -27.11 & 0.10 & -26.96 \\
region-slice-seed42& 400 & 10 & \textbf{0.814} & \textbf{0.443} & -26.92 & 0.33 & -26.96 \\
& 400 & 20 & 1.023 & 0.650 & -26.99 & 0.30 & -26.96 \\
de-mix-seed42& 400 & 10 & \textbf{0.985} & 0.561 & -27.79 & 0.41 & -26.96 \\
& 400 & 20 & 1.079 & 0.690 & -27.32 & 0.44 & -26.96 \\
& 400 & 40 & 1.237 & 0.804 & -26.56 & 0.38 & -26.96 \\
& 400 & 80 & 1.265 & 0.829 & -26.55 & 0.35 & -26.96 \\
& 400 & 160 & 1.267 & 0.822 & -27.03 & 0.47 & -26.96 \\
region-rwm-seed42& 400 & 10 & \textbf{0.082} & \textbf{0.027} & -42.40 & 0.29 & -26.96 \\
& 400 & 20 & \textbf{0.261} & \textbf{0.029} & -32.37 & 0.38 & -26.96 \\
& 400 & 40 & \textbf{0.234} & \textbf{0.042} & -33.25 & 0.27 & -26.96 \\
& 400 & 80 & \textbf{0.330} & \textbf{0.075} & -29.13 & 0.37 & -26.96 \\
& 400 & 160 & \textbf{0.455} & \textbf{0.108} & -27.87 & 0.47 & -26.96 \\
& 400 & 320 & \textbf{0.551} & \textbf{0.131} & -26.51 & 0.47 & -26.96 \\
& 400 & 640 & \textbf{0.704} & \textbf{0.172} & -27.75 & 0.42 & -26.96 \\
\hline
\end{tabular}

\end{table}

\begin{table}
    \centering
    \caption{Summary of runs, continued, for the funnel with d=10 (corrfunnelN-10d).}
    \label{tab:summaryfunnelseed}
\begin{tabular}{lrrrrrrrl}
\hline
Variant & $K$ & Steps & <RJD> & $f(\mathrm{RJD}>1)$ & $\ln Z$ & $\sigma_{\ln Z}$ & $\ln Z_\mathrm{true}$ \\
\hline
cube-slice-seed1& 400 & 10 & \textbf{0.911} & \textbf{0.477} & -27.55 & 0.46 & -26.96 \\
& 400 & 20 & 1.099 & 0.668 & -26.93 & 0.33 & -26.96 \\
& 400 & 40 & 1.173 & 0.742 & -26.56 & 0.45 & -26.96 \\
& 400 & 80 & 1.213 & 0.779 & -26.55 & 0.44 & -26.96 \\
& 400 & 160 & 1.222 & 0.790 & -27.13 & 0.47 & -26.96 \\
cube-slice-seed2& 400 & 10 & \textbf{0.914} & \textbf{0.473} & -26.79 & 0.35 & -26.96 \\
& 400 & 20 & 1.070 & 0.645 & -26.57 & 0.32 & -26.96 \\
& 400 & 40 & 1.153 & 0.723 & -26.64 & 0.29 & -26.96 \\
cube-slice-seed3& 400 & 10 & \textbf{0.907} & \textbf{0.462} & -26.52 & 0.36 & -26.96 \\
& 400 & 20 & 1.083 & 0.658 & -27.42 & 0.34 & -26.96 \\
& 400 & 40 & 1.180 & 0.743 & -27.36 & 0.26 & -26.96 \\
& 400 & 80 & 1.211 & 0.780 & -26.92 & 0.39 & -26.96 \\
& 400 & 160 & 1.237 & 0.798 & -26.93 & 0.37 & -26.96 \\
cube-slice-seed4& 400 & 10 & \textbf{0.903} & \textbf{0.468} & -27.89 & 0.38 & -26.96 \\
& 400 & 20 & 1.086 & 0.652 & -26.38 & 0.42 & -26.96 \\
& 400 & 40 & 1.186 & 0.752 & -26.98 & 0.31 & -26.96 \\
& 400 & 80 & 1.218 & 0.786 & -27.05 & 0.28 & -26.96 \\
& 400 & 160 & 1.211 & 0.780 & -27.06 & 0.44 & -26.96 \\
& 400 & 320 & 1.218 & 0.787 & -26.57 & 0.39 & -26.96 \\
\hline
\end{tabular}

\end{table}

\end{document}